\documentclass[runningheads,a4paper]{llncs}

\usepackage{amssymb}
\usepackage{graphicx}
\usepackage{color}
\usepackage{xcolor}
\usepackage{multirow}
\usepackage{csquotes}
\usepackage{multicol}
\usepackage{subcaption}
\usepackage{tikz}
\usepackage{comment}
\usetikzlibrary{tikzmark}
\usepackage{colortbl}
\usepackage{array}
\newcolumntype{L}[1]{>{\raggedright\let\newline\\\arraybackslash\hspace{0pt}}m{#1}}
\newcolumntype{C}[1]{>{\centering\let\newline\\\arraybackslash\hspace{0pt}}m{#1}}
\newcolumntype{R}[1]{>{\raggedleft\let\newline\\\arraybackslash\hspace{0pt}}m{#1}}

\usepackage{url}

\urldef{\mailsa}\path|{m.a.m.soliman, p.avgeriou}@rug.nl|   
\urldef{\mailga}\path|{mgalster}@ieee.org| 
\newcommand{\keywords}[1]{\par\addvspace\baselineskip
\noindent\keywordname\enspace\ignorespaces#1}

\begin{document}
\bibliographystyle{splncs}
\mainmatter

\title{An Exploratory Study on Architectural Knowledge in Issue Tracking Systems}
\titlerunning{An Exploratory Study on Architectural Knowledge in Issue Tracking Systems}
\author{Mohamed Soliman\thanks{Corresponding author. \\ We would like to thank ITEA3 and RVO for their support under grant agreement No. 17038 VISDOM (https://visdom-project.github.io/website).}\inst{1}, Matthias Galster\inst{2} \and Paris Avgeriou\inst{1}}

\institute{University of Groningen, Netherlands\\
\mailsa
\and
University of Canterbury, New Zealand\\
\mailga}

\maketitle

\begin{abstract}
Software developers use issue trackers (e.g. Jira) to manage defects, bugs, tasks, change requests, etc. In this paper we explore (a) how architectural knowledge concepts (e.g. architectural component behavior, contextual constraints) are textually represented in issues (e.g. as adjectives), (b) which architectural knowledge concepts commonly occur in issues, and (c) which architectural knowledge concepts appear together. We analyzed  issues in the Jira issue trackers of three large Apache projects. To identify ``architecturally relevant'' issues, we linked issues to architecturally relevant source code changes in  the studied systems. We then developed a code book by manually labeling a subset of issues. After reaching conceptual saturation, we coded remaining issues. Our findings support empirically-grounded search tools to identify architectural knowledge concepts in issues for future reuse.

\end{abstract}

\section{Introduction}

Architectural design decisions (ADDs) about software components, their dependencies and behavior are one of the most significant types of ADDs  \cite{BassBook2012} made by practitioners \cite{Weinreich2013}. For example, an enterprise application could have multiple ADDs regarding the components in each layer and dependencies between them. In the Architectural Knowledge (AK) ontology of Kruchten et al. ~\cite{Kruchten2006}, ADDs related to component design are identified as \textit{structural ADDs} (SADDs).


Making SADDs, requires AK \cite{Kruchten2006} about important quality attributes and their trade-offs, and about instantiating architectural solutions (e.g. patterns or tactics). Without the required AK, software engineers might make uncertain and risky assumptions about the ADDs. However, this AK is mostly acquired through experience with multiple different projects within the same domain \cite{vanVliet2016DecisionArchitecture}. If one does not have such experience, one needs to be able to search and locate the pertinent AK.


One potential source of AK is issue tracking systems (e.g. Jira): previous research has shown that software engineers share some AK (e.g. previously made ADDs) in issues \cite{Bhat2017AutomaticApproach,Shahbazian2018RecoveringDecisions}. Developers create issues to discuss and manage defects, tasks, change requests, etc. 
However, similar to other sources of AK (e.g. developer communities \cite{SolimanICSA2017,Wicsa2016} or simply Google searching \cite{Soliman2021ExploringKnowledge}), it is challenging to manually recognize and re-use AK within issue tracking systems. First, the majority of issues do not discuss architectural problems \cite{Bhat2017AutomaticApproach};  instead, they focus on detailed development problems (e.g. bug fixing or testing). 
Second, text in issues is not explicitly structured and classified; rather, AK is represented as unstructured text within an issue's description, comments and attachments.

Recently, there has been research work on automatic mining of AK from issues: there are studies that identify types of ADDs \cite{Bhat2017AutomaticApproach} and types of architectural issues \cite{Shahbazian2018RecoveringDecisions}. However, they do not explore  \textbf{AK concepts} (i.e. conceptual elements that describe and characterize AK), such as \emph{types of architectural solutions} (e.g. components behavior and tactics) \cite{Jansen2005,Zimmermann2009}, \emph{decision factors} like constraints, and \emph{decision rationale} \cite{Tang2009SoftwareReasoning} like assumptions, benefits and drawbacks of solutions \cite{SolimanWicsa2015}. Moreover, current studies \cite{Bhat2017AutomaticApproach} limit their analysis to issue descriptions which are only a small part of the whole issue (often the shortest) without exploring comments in issues or attachments. Finally, current studies do not explore how AK concepts are textually represented (e.g. using adjectives or explicit keywords) in issues. These three limitations make it nearly impossible to determine AK concepts in architectural issues, and prevent approaches to find, capture and re-use AK from issue tracking systems.

We contribute in addressing these shortcomings by \textit{exploring the different AK concepts for making SADDs and their representation in architectural issues} (see research questions in Section~\ref{sec:studydesign}). We look at the entire issue, instead of only its description.
Achieving this goal 
supports automating AK capturing approaches with concrete representations for AK concepts. Moreover, it supports determining the most suitable scenarios for re-using AK from issue tracking systems (e.g. when searching for alternative solutions or comparing solutions).

To achieve this goal, we conduct a case study on three large Apache projects. We analyze \emph{architecturally relevant} issues from the projects' issue tracker. The issues are first identified by static analysis of the source code of the projects, and then the textual contents of issues are verified as architecturally relevant and analyzed to explore their contained AK concepts (see Section \ref{sec:studydesign}). In summary, our study leads to the following contributions:
\begin{itemize}
\item A corpus of 98 architectural issues, and 3,937 annotations for AK concepts. This helps future research, e.g. using machine learning to capture AK.
\item Common textual variants for each AK concept in architectural issues. This is useful to identify and search for AK concepts in issue tracking systems.
\item A list of the most discussed AK concepts in issue tracking systems. This supports identifying scenarios in which AK from issues can be re-used.
\item Common co-occurrences of different AK concepts in architectural issues. This supports capturing relationships between AK concepts from these issues.
\end{itemize}
The paper is structured as follows: In Section \ref{sec:background}, we provide a background on relevant AK concepts. In Section \ref{sec:studydesign}, we explain our research questions and steps. We then present our results in Sections \ref{sec:textualpatterns}, \ref{sec:AKConceptsNumbers} and \ref{sec:AKConceptsCooc}, which are subsequently discussed in Section \ref{sec:discussion}. The threats to validity and related work are discussed in Sections \ref{sec:threats} and \ref{sec:related}, while the paper is concluded in Section \ref{sec:conclusion}.
\vspace{-10.0pt}
\section{Background - Architectural Knowledge Concepts}
\label{sec:background}
\vspace{-10.0pt}
In this section, we give an overview of AK concepts in the literature, which are relevant to this study. We consider AK concepts from different studies, because there is no comprehensive ontology with all AK concepts. Each concept is represented by an abbreviation, that is used in the rest of the paper.

Software engineers consider different \emph{decision factors} \cite{BassBook2012,SolimanWicsa2015}:
\begin{itemize}
\item \textit{Requirements and constraints} (REQ) could be \textit{quality attribute requirements}, such as performance, maintainability, security \cite{BassBook2012}, \textit{user functional requirements}, such as use cases and user stories, or \textit{contextual constraints} such as external systems or constraints from managers  \cite{BassBook2012}.
\item \textit{Architecture of existing system} (EXA) may constrain new ADDs \cite{Gerdes2014CombiningEvolution}.
\item \textit{Quality issues of existing system} (EXQ) can involve \textit{run-time quality issues} (e.g. performance issues) or \textit{technical debt items} (e.g. architectural smells). While REQ may represent the target value for a quality requirement, EXQ is the current value that needs improvement (e.g. security vulnerabilities).
\end{itemize}

Furthermore, ADDs require deciding on one or more \emph{architectural solutions} \cite{Zimmermann2009}. These could have several types, such as:
\begin{itemize}
\item \textit{Architectural component behavior} (CB) describes the behavior of an architecture component, including the implemented logic and complexity~\cite{SolimanICSA2017}.
\item \textit{Architectural configuration} (CONF) describes the dependencies of components~\cite{Medvidovic2000,SolimanICSA2017}.
\item \textit{Architectural tactics} (TAC) address specific quality attributes, for example, caching data (tactic) improves performance  \cite{BassBook2012}.
\end{itemize}

Finally, ADDs should be based on a certain \emph{rationale} \cite{Tang2009SoftwareReasoning} (i.e. the reason for selecting architectural solutions). This involves several AK concepts:
\begin{itemize}
\item \textit{Architectural solution benefits and drawbacks} (ABD) describe strengths and weaknesses for certain architectural solutions \cite{SolimanWicsa2015}.
\item \textit{Assumptions} (ASSUM) capture facts which are assumed without proof when deciding on an architectural solution \cite{Yang2017AnFramework}.
\item \textit{Trade-offs} (TRO) describe balanced analysis of what is an appropriate option after prioritizing and weighing different design options \cite{Tang2009SoftwareReasoning}.
\item \textit{Risks} (RIS) capture considerations about uncertainties of design options \cite{Tang2009SoftwareReasoning}.
\end{itemize}






\section{Study Design}
\label{sec:studydesign}
\vspace{-10.0pt}
\subsection{Research Questions}
To achieve our goal, we ask the following research questions:
\begin{itemize}
\item \textit{(\textbf{RQ1}) How are AK concepts textually represented within architectural issues that discuss SADDs?}
Since AK concepts (Section \ref{sec:background}) are high-level conceptual entities, their textual representation could come in multiple different forms. Determining how AK concepts are actually represented (e.g. using certain keywords or adjectives) in architectural issues 
can support improving the accuracy of automatically identifying AK concepts.
\item \textit{(\textbf{RQ2}) Which AK concepts are commonly used by practitioners within architectural issues to discuss SADDs?} While researchers have empirically explored several AK concepts (Section \ref{sec:background}) and how they are used by practitioners (e.g. on Stack Overflow \cite{SolimanICSA2017}), it is unknown which AK concepts are shared in issue tracking systems to make SADDs. Identifying AK concepts, which are commonly discussed in issues, is useful in determining scenarios to re-use the AK in architectural issues.


\item \textit{(\textbf{RQ3}) Which AK concepts co-occur with each other when discussing SADDs in architectural issues?} The discussion of SADDs in an issue, often does not pertain to a single AK concept, but may involve multiple related AK concepts. Conceptual relationships between AK concepts in ADDs have been discussed in literature (e.g. \cite{Zimmermann2009,Jansen2005}). However, it is unknown how practitioners use different AK concepts together to discuss SADDs in architectural issues. 
Determining common co-occurrences between AK concepts provides contextual relationships between the different AK concepts. These are important to support capturing related AK concepts from architectural issues.


\end{itemize}
\vspace{-20.0pt}
\subsection{Research Process}
\vspace{-30.0pt}
\begin{figure*}
\setlength{\belowcaptionskip}{-10pt}
	\centering
		\includegraphics[scale=0.54]{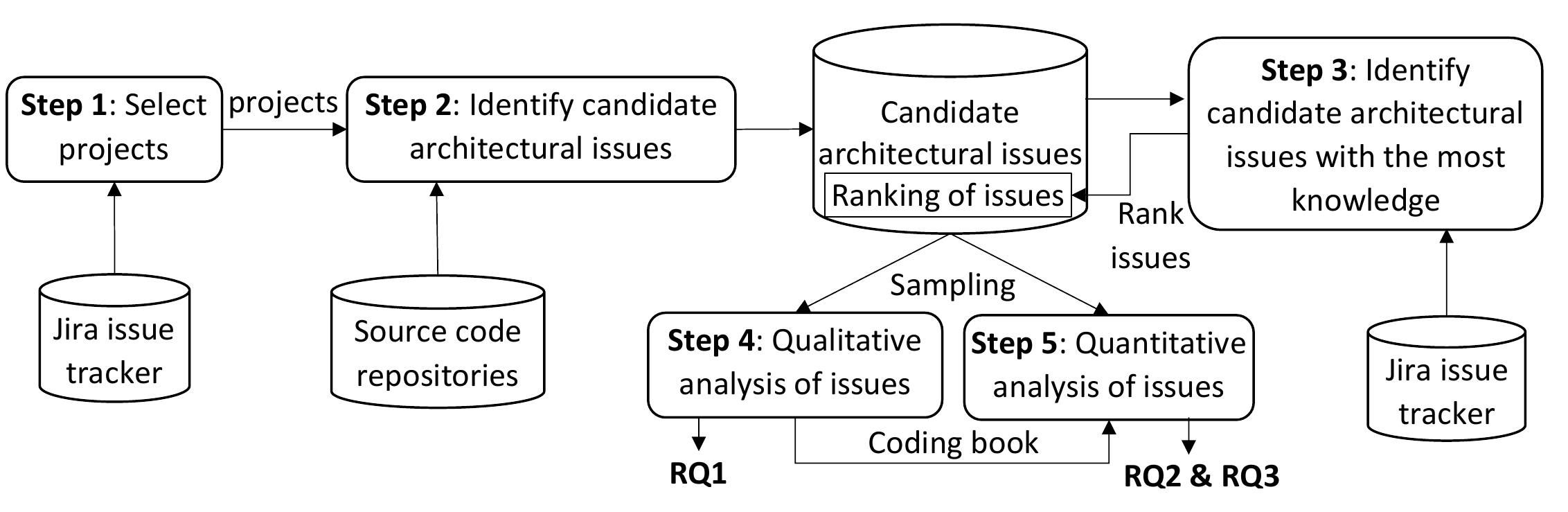}       
		\caption{Research steps}
	\label{fig:researchprocess}
\end{figure*}
We followed five steps (see Fig. \ref{fig:researchprocess}) as explained in the following sub-sections.

\subsubsection{Step 1 - Select projects:}
We selected Apache Java projects, similarly to previous research on AK in issues \cite{Bhat2017AutomaticApproach,Shahbazian2018RecoveringDecisions},
and based on the following criteria:
\begin{enumerate}
\item \textit{Larger than 200 KLOC}: This is to ensure sufficiently large projects with a sufficient number of architectural issues. 
\item \textit{Use of Jira}: Jira is a commonly used issue tracking system. It also provides useful features for managing issues (e.g. an API to download issues).
\item \textit{Traceability between code commits and issues}: To identify candidate architectural issues (Step 2) based on links between code commits and issues, we require that issue IDs are used in commit messages of the code repository.
\item \textit{Sufficient knowledge within issues}: Not all Apache projects discuss decisions within issues. Some projects use other communication methods (e.g. mailing lists). 
Thus, we have to ensure that Jira not only lists tasks, but also communications. We calculated the average number of comments within issues for each project, and the percentage of issues with no comments (i.e. no discussions in issues). We then ranked projects (available online\footnote{\label{onlinematerial}github.com/m-a-m-s/ECSA-2021}) to identify projects with the most issues and the most comments within issues.
\end{enumerate}
We selected the top three Apache projects from the resulted ranking in the fourth criterion (with the most issues and comments in issues): Apache Hadoop, Cassandra, and Tajo.
\vspace{-10.0pt}
\subsubsection{Step 2: Identify candidate architectural issues}
\label{sec:identifyissues}
Some issues involve architectural changes \cite{Shahbazian2018RecoveringDecisions}. However, most issues in tracking systems do not trigger architectural changes \cite{Shahbazian2018RecoveringDecisions}, because they involve small changes within components (e.g. small bug fixes). Thus, we need to identify the issues which trigger architectural changes. This has been addressed previously in different ways:
\begin{enumerate}
\item \textit{Top-down}: Bhat et al. \cite{Bhat2017AutomaticApproach} manually analyze issues to determine if they involve discussions on architectural decisions. This requires significant effort, because of the sheer volume of issues (e.g. Hadoop has more than 50,000 issues). Thus, the selected architectural issues based on this approach might not be representative for the most significant architectural issues.
\item \textit{Bottom-up}: Shahbazian et al. \cite{Shahbazian2018RecoveringDecisions} analyze source code, and construct dependency graphs for consecutive versions of a project. Differences between dependency graphs are compared to assess if changes within each version are architectural. The source code versions are then traced back to issues to identify architectural issues. The approach can effectively identify candidates for architectural issues. However, inaccuracy in the assessment of architectural changes might miss architectural issues or identify false positives.
\end{enumerate}
Similar to Shahbazian et al. \cite{Shahbazian2018RecoveringDecisions}, we follow a bottom-up approach to identify candidate architectural issues with the most architectural changes. However, we further filtered issues with the most AK and to reduce false positives. In detail, we followed three steps to identify candidate architectural issues:
\begin{enumerate}
\item \textit{Construct dependency graphs from source code}: We used Arcan's system re-constructor \cite{Fontana2017Arcan:Detection} to create dependency graphs (one graph for each version) between classes and packages of the three projects.
\item \textit{Estimate architectural changes}: We compared the dependency graph of each version with the dependency graph of its preceding version and determined added or removed Java packages, added or removed dependencies between packages and changes in the allocations of classes to packages. We chose Java packages as architectural components, because they have been explicitly designed by the developers. Moreover, packages are at a higher abstraction level than classes. Thus, changes at package level are likely more architectural than changes at class level. To compare the architecture of two consecutive versions we calculated the Architecture-to-Architecture (a2a) metric \cite{Behnamghader2017ASystems}:
\begin{enumerate}
\item Calculate \textit{Minimum Transform Operation (MTO)} for two consecutive versions in a repository. The MTO between two versions is the sum of added packages, removed packages, added dependencies, removed dependencies and the number of re-allocated classes between packages.
\item Calculate the a2a metric \cite{Behnamghader2017ASystems} for each version in a repository as the percentage of MTO for a certain version compared to the size of the architecture (i.e. total number of packages, dependencies and classes).
\end{enumerate}
The a2a metric has been previously used by Shahbazian et al. \cite{Shahbazian2018RecoveringDecisions}, and can provide a reliable estimation for the size of architectural changes.
\item \textit{Filter and rank architectural issues}: We identified versions in source code repositories with a2a \textgreater 0 and identified related issues. Issues have been discovered by following the traceability links (Jira issue ID's, e.g. TAJO-88, used in commit messages on GitHub) between versions in GitHub and issues in Jira. We have identified 2,575 candidate architectural issues (from over 28,000 issues in the three projects), which are responsible for possible architectural changes. The candidate architectural issues are shared online.
\end{enumerate}
\vspace{-15.0pt}
\subsubsection{Step 3: Identify candidate architectural issues with most AK}
\label{sec:selectissues}
From the 2575 candidate architectural issues, we need to select those that contain actual AK and of sufficient quantity. Thus, we identified the issues with the most AK based on three steps:
\begin{enumerate}
\item \textit{Identify parent issues and sub-task issues}: For all candidate architectural issues (Step 2), we identified their parent issue (if the issue is a sub-task), and their sub-task issues (if the issue is a parent issue).
\item \textit{Estimate amount of AK per issue}: For all candidate architectural issues (Step 2), their parents and sub-tasks, we counted the number of words, considering issue description, comments and attachments (e.g. pdf documents). 
\item \textit{Rank candidate architectural issues}: We ranked issues per project according to their architectural significance (based on the a2a metric from Step 2) and the amount of knowledge per issue (based on the number of words per issue).
The ranked issues present the population from which a sample of issues are selected for analysis in Step 4 and another sample in Step 5. 
\end{enumerate}
\vspace{-15.0pt}
\subsubsection{Step 4: Qualitative analysis of AK concepts in issues}
To answer RQ1, we analyzed a sample of issues with the most AK from the ranked list of candidate architectural issues (from Step 3) qualitatively.
We followed deductive category assignment content analysis as defined by Mayring \cite{Mayring2000}:
\begin{enumerate}
\item \textit{Identify AK concepts from literature}: To identify AK concepts, we needed a category system based on existing literature. Thus, we reviewed literature in the field of AK and identified AK concepts (Section \ref{sec:background}). The AK concepts and their definition were the starting point for the coding book.
\item \textit{Preliminary coding to create initial coding book}: The first two authors annotated independently a sample of architectural issues with the AK concepts identified from literature. Before annotating an issue, we manually validated that an issue is actually an architectural issue, and that it contains AK. During annotation, we followed two \textit{annotation rules}:
\begin{itemize}
\item We annotated AK concepts (Section \ref{sec:background}) as clauses or sentences or paragraphs, because AK concepts do not appear as single words.
\item We ignored all textual segments with no relationship to AK concepts, such as code examples and test executions.
\end{itemize}
For each issue, the first two authors compared annotations and discussed differences in meetings. After each meeting, we refined the coding book with concrete definitions and examples for each AK concept. Our aim was to operationalize abstract AK concepts (from literature) into concrete definitions (in the issues). After several iterations and annotating 20 randomly selected issues from the sample with more than 300 annotations, we reached theoretical saturation (i.e. no new AK concepts appeared in issues).
\item \textit{Identify textual variants of AK concepts in issues}: Based on the preliminary coding, we identified textual variants for each AK concept to answer RQ1. For example, benefits and drawbacks of solutions are expressed explicitly using certain keywords (e.g. advantages) or adjectives. Moreover, we added the textual variants for each AK concept to the coding book. This supports other researchers to annotate the same AK concepts reliably. The most common textual variants for each AK concept are presented in Section \ref{sec:textualpatterns}.
\end{enumerate}
\vspace{-15.0pt}
\subsubsection{Step 5: Quantitative analysis of AK concepts in issues}
To answer RQ2 and RQ3, we performed the following steps:
\begin{table}[h]
\setlength{\belowcaptionskip}{-25pt}
\centering
\begin{tabular}{ L{3cm} L{9cm} }
\hline
\multicolumn{1}{C{3cm}}{\textbf{AK concepts and variants}} & \multicolumn{1}{C{9cm}}{\textbf{Description and examples}} \\ 
\rowcolor{lightgray}  \multicolumn{2}{l}{\tikzmark{QA}\textit{Quality attribute as one type of requirements} (REQ)}   \\ 
\hspace{0.3cm}  \tikzmark{EXP}Explicitly &   Uses common quality attribute-related terms like ``extensibility'' or ``performance''. For example: \textit{``This improves the code \textbf{readability} and \textbf{maintainability}''} [TAJO-121] \\ \cline{2-2}
\setlength{\leftskip}{0.3cm} \tikzmark{IMP}Implicitly using adjectives &  Describes the quality of certain system or component using adjectives.
For example ``\textit{For the sake of \textbf{efficient} join order enumeration,...}'' [TAJO-229] could point to performance. \\ 
\rowcolor{lightgray} 
\multicolumn{2}{l}{\tikzmark{EXQ}\textit{Existing system quality} (EXQ)}   \\ 
\setlength{\leftskip}{0.3cm} \tikzmark{NEG}Negation of quality &  Refers to a component of a system with negation keywords and adjectives.
For example, ``\textit{It is rather \textbf{complicated} and \textbf{does not} guarantee data recoverability}'' [HADOOP-702]  \\ \cline{2-2} 
\setlength{\leftskip}{0.3cm} \tikzmark{QUISS}Explicit quality issues &  Describes well-known quality issues using their terms explicitly 
For example, ``\textit{The dependencies between them should be enforced to avoid \textbf{cyclic dependencies}. At present they all have \textbf{dependencies on each other}}'' [HADOOP-3750] \\
\hline
\tikz[remember picture] \foreach \i in {EXP,IMP} \draw[overlay] (pic cs:QA) |- ([yshift=1.5mm]pic cs:\i);
\tikz[remember picture] \foreach \i in {NEG,QUISS} \draw[overlay] (pic cs:EXQ) |- ([yshift=1.5mm]pic cs:\i);
\end{tabular}
\caption{Variants of most common \textbf{decision factors} AK concepts}
\label{tab:decisionfactorsAK}
\end{table}
\begin{enumerate}
\item \textit{Expand annotations to ensure statistical significance}: 
We provided the coding book (from the previous two steps) to two independent researchers to annotate AK concepts in selected architectural issues from the population of candidate architectural issues. We randomly selected issues proportional to their ranking in the list of candidate issues (from Step 3). 
This method of sampling supports selecting issues with the most AK to ensure exploring AK concepts.
The first author explained the coding book to the two researchers who followed the same annotation rules as in the preliminary coding. In several iterations, a sample of issues were independently annotated by the first author to ensure agreement. Disagreements were discussed to ensure understanding of the coding book. Before annotating an issue, the first author checked its architectural relevance.

To answer RQ2 and RQ3, we need to ensure that the number of annotations is sufficient for higher confidence level and lower error margin. However, some issues involve lots of discussions and thus can involve hundreds of annotations, while others are nearly empty. Thus, we had to first estimate the possible number of annotations in our population of candidate architectural issues. We did this by dividing the total number of words ($\approx$ 20 millions words) in all candidate architectural issues (from Step 3) by the average size of each annotation ($\approx$ 25 words, see Section \ref{sec:AKConceptsNumbers}). Thus, the estimated number of annotations in the whole population is $\approx$ 800,000 annotations. Accordingly, we created a statistically significant sample size \cite{NeuendorfTheGuidebook} of 3,937 annotations with 95\% confidence level and 1.6\% error margin. The annotations are created from 98 architectural issues with the most architectural significance and most AK inside them.

\item \textit{Analyze annotations to answer research questions}: For RQ2 we counted the number of annotations for each AK concept. To determine their sizes, we counted words after removing stop-words. For RQ3 we counted the number of co-occurrences of annotations (for each AK concept) which occur together in either an issue description or comment. We then tested the significance of each co-occurrence of AK concepts using a $\tilde{\chi}^2$ test \cite{PearsonSquare}. For example, for the AK concepts \emph{Benefits and drawbacks (ABD)} and \emph{Component behavior (CB)}, we considered frequencies for the following four situations: 1) Text annotated as ABD co-occur with annotations for CB. 2) Text annotated as ABD co-occur with annotations other than CB. 3) Text annotated with AK concepts other than ABD co-occur with annotations for CB. 4) Text annotated with AK concept other than ABD co-occur with annotations other than CB. We excluded co-occurrences with $\tilde{\chi}^2 < 10$ to ensure that all co-occurrences were statistically significant at p-value \textless 0.05. The significant co-occurrences between AK concepts are presented in Section \ref{sec:AKConceptsCooc}.

\end{enumerate}
\vspace{-20.0pt}
\section{RQ1: Representation of AK Concepts in Issues}
\label{sec:textualpatterns}
\begin{table}[h]
\setlength{\belowcaptionskip}{-20pt}
\centering
\begin{tabular}{ L{3cm} L{9cm} }
\hline
\multicolumn{1}{C{3cm}}{\textbf{AK concepts and variants}} & \multicolumn{1}{C{9cm}}{\textbf{Description and examples}} \\ 
\rowcolor{lightgray}  \multicolumn{2}{l}{\tikzmark{BD}\textit{Benefits and drawbacks} (ABD)}   \\ 
\hspace{0.3cm}  \tikzmark{EXPBD}Explicitly &  Using terms like ``advantages'', ''limitations', etc. For example ''\textit{Keeping things config-file based has two \textbf{drawbacks}:...}'' [CASSANDRA-44]. \\ \cline{2-2}
\setlength{\leftskip}{0.3cm} \tikzmark{ADJBD} Using adjectives & 
Adjectives could be generic like ''good'', ''ugly'', etc. For example, ``\textit{it would be a \textbf{fragile} solution to the identified problem}'' [HADOOP-1053]. It could also be more related to special quality attribute. For example ''\textit{The serialization mechanism proposed so far is...so \textbf{general}}'' [HADOOP-1986]. \\ \cline{2-2}
\setlength{\leftskip}{0.3cm} \tikzmark{MESBD} Using quality measurement & 
Expressing special quality measurements. For example, ``\textit{We can do group by aggregations on \textbf{billions of rows with only a few milliseconds}}'' [TAJO-283]. \\ \cline{2-2}
\setlength{\leftskip}{0.3cm} \tikzmark{PROBBD} Problems in a solution & Problems which are a consequence from using a particular solution. For example ``\textit{multiget-within-a-single-row still \textbf{has all the problems} of multiget-across-rows...it doesn’t parallelize across machines}'' [CASSANDRA-2710]. \\
\rowcolor{lightgray} 
\multicolumn{2}{l}{\tikzmark{ASS}\textit{Assumptions} (ASSUM)}   \\ 
\hspace{0.3cm}  \tikzmark{ASSEXP}Explicitly &  Explicit references to assumptions, e.g., the word ``assumption'' or synonyms. 
For example, ``\textit{\textbf{Assume} that Jobs and interfering cache updates won’t occur concurrently}'' [HADOOP-288]  \\ \cline{2-2} 
\setlength{\leftskip}{0.3cm} \tikzmark{ASSIMP}Using uncertainty terms & Uncertain and vague terms, such as ``I think'', ``it might`''. For example, ``\textit{\textbf{I think}, saving values would limit fexibility of the cache  interface...}'' [CASSANDRA-3143] \\ 
\hline
\tikz[remember picture] \foreach \i in {EXPBD,ADJBD,MESBD,PROBBD} \draw[overlay] (pic cs:BD) |- ([yshift=1.5mm]pic cs:\i);
\tikz[remember picture] \foreach \i in {ASSEXP,ASSIMP} \draw[overlay] (pic cs:ASS) |- ([yshift=1.5mm]pic cs:\i);
\end{tabular}
\caption{Variants of most common \textbf{rationale} AK concepts}
\label{tab:rationaleAK}
\end{table}
Based on the analysis of architectural issues (see Section \ref{sec:studydesign}), we identified the representation of AK concepts in terms of common textual variants for each AK concept. The most frequent textual variants for the most common AK concepts (see Section \ref{sec:AKConceptsNumbers}) are explained in Tables \ref{tab:decisionfactorsAK}, \ref{tab:rationaleAK}, and \ref{tab:architecturalsolutionsAK}. Further variants are provided online. Some of the decision factors (Table \ref{tab:decisionfactorsAK}) and rationale AK concepts (Table \ref{tab:rationaleAK}) occur explicitly or implicitly. The explicit variants can be easily identified, because they depend on the occurrence of certain keywords (e.g. ``performance'' for quality attributes or ``advantage'' for benefits and drawbacks). Implicit variants are more difficult to distinguish, since they rely on a combinations of words, which must appear together in a certain context to deliver the meaning of the AK concept. For example, both quality attributes and benefits and drawbacks could be expressed using adjectives. However, benefits and drawbacks are expressed in combination with a certain architectural solution, while quality attributes are expressed in relation to certain requirements. This presents a challenge to accurately identify AK concepts.

We also observed domain-specific terms in some of the variants to describe the architecture solutions (Table \ref{tab:architecturalsolutionsAK}). For example, optimization of queries is a core functionality in Apache Tajo, and the term ``optimizer'' refers to an architectural component. This might not be the case in other systems, where an optimizer might be a tool to improve source code. Thus, AK in architectural issues depends strongly on the domain and context. This poses a challenge in finding AK in issue trackers, which describes the architecture components of a system.
\vspace{-15pt}
\begin{table}[h]
\setlength{\belowcaptionskip}{-20pt}
\centering
\begin{tabular}{ L{3cm} L{9cm} }
\hline
\multicolumn{1}{C{3cm}}{\textbf{AK concepts and variants}} & \multicolumn{1}{C{9cm}}{\textbf{Description and examples}} \\ 
\rowcolor{lightgray}  \multicolumn{2}{l}{\tikzmark{CB}\textit{Architectural component behavior} (CB)}   \\ 
\hspace{0.3cm}  \tikzmark{APP}Approach &   Describes the main approach (e.g. algorithm) on which behavior of component is based. For example ``\textit{So the \textbf{approach} I propose is...to iterate through the key space on a per-CF basis, compute a hash...}'' [CASSANDRA-193]. \\ \cline{2-2}
\setlength{\leftskip}{0.3cm} \tikzmark{SUB}Sub-components & Describes sub-components (e.g. interfaces or data structures) which implement the component's behavior, 
e.g. ``\textit{This optimizer will \textbf{provide the interfaces} for join enumeration algorithms and rewrite rules}'' [TAJO-24 attachment]. \\ 
\rowcolor{lightgray} 
\multicolumn{2}{l}{\tikzmark{CONF}\textit{Architectural design configuration} (CONF)}   \\ 
\setlength{\leftskip}{0.3cm}  \tikzmark{STAT}Static dependencies &  Dependencies between components independent of their sequence. The dependencies can use connector verbs such as ``access'' and ``obtain'' \cite{SolimanICSA2017}.
For example ``\textit{a management tool to \textbf{contact} every node via JMX}'' [CASSANDRA-44]. \\ \cline{2-2} 
\setlength{\leftskip}{0.3cm}  \tikzmark{DYN}Dynamic dependencies &  Sequence of interactions between components, for example ``\textit{YARN mode \textbf{1.} TajoClient request query to TajoMaster. \textbf{2.} YarnRMClient request QueryMaster(YARN Application Master)...}'' [TAJO-88 attachment] \\ 
\hline
\tikz[remember picture] \foreach \i in {APP,SUB} \draw[overlay] (pic cs:CB) |- ([yshift=1.5mm]pic cs:\i);
\tikz[remember picture] \foreach \i in {STAT,DYN} \draw[overlay] (pic cs:CONF) |- ([yshift=1.5mm]pic cs:\i);
\end{tabular}
\caption{Variants of most common \textbf{architectural solutions} AK concepts}
\label{tab:architecturalsolutionsAK}
\end{table}
\vspace{10pt}
\section{RQ2: Prominent AK Concepts in Issues}
\label{sec:AKConceptsNumbers}
\vspace{-20.0pt}
\begin{figure}[ht]
\setlength{\belowcaptionskip}{-10pt}
\begin{subfigure}{.5\textwidth}
  \centering
  \includegraphics[width=0.75\linewidth]{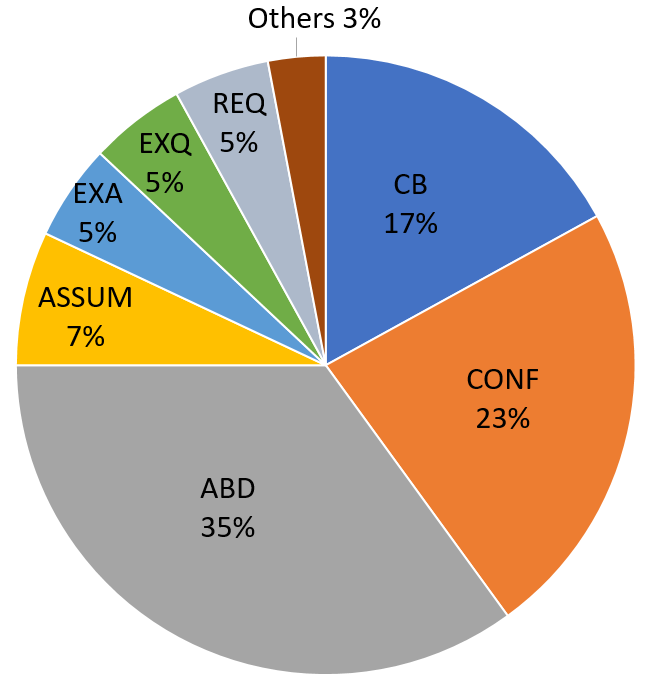} 
  \caption{\% of annotations for each AK concept}
  \label{fig:sub-first}
\end{subfigure}
\begin{subfigure}{.5\textwidth}
  \centering
  \includegraphics[width=1.00\linewidth]{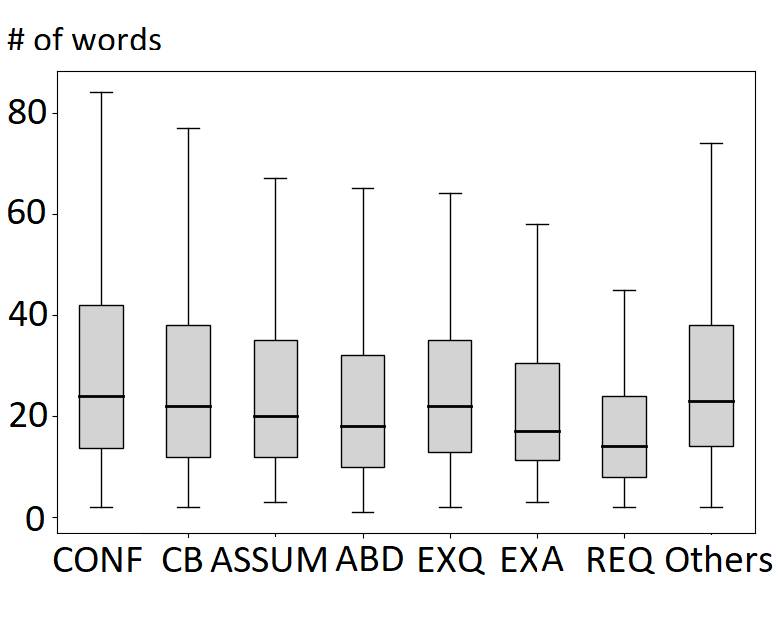}  
  \caption{Size of annotations for each AK concept}
  \label{fig:sub-second}
\end{subfigure}
\caption{The amount and size of annotations for each AK concept}
\label{fig:AKConcepts}
\end{figure}
Fig. \ref{fig:AKConcepts} shows the percentages of annotations related to each AK concept, as well as their size (number of words). 
There are no significant differences in the percentages or sizes of annotations among issue descriptions and issue comments.
Issues tend to include AK regarding proposed architectural solutions ($\approx$ 40\%) and their rationale  (\textgreater 40\%) more frequently than  decision factors ($\approx$ 15\%). AK regarding architectural solutions focuses mainly on components behavior (CB) and architectural configurations (CONF), which align with the scope of this paper to explore SADDs. The benefits and drawbacks (ABD) of architectural solutions dominate the rationale of ADDs in architectural issues, followed by Assumptions (ASSUM). However, trade-offs and risks are rarely shared in architectural issues. Also, descriptions of architectural solutions (i.e. CONF and CB) tend to be larger than their rationale (i.e. ABD and ASSUM). This means that developers describe their architectural solutions extensively, but provide brief justifications for decisions.

Most decision factors are about the architecture (EXA) and quality issues (EXQ) of an existing system (both $\approx$ 10\%). Discussions on requirements and constraints (REQ) are rare ($\approx$ 5\%). Also, REQ annotations are the shortest, which indicates limited discussions about architectural significant requirements.
\vspace{-20pt}
\section{RQ3: Significant Co-occurrences between AK Concepts}
\label{sec:AKConceptsCooc}
\vspace{-20.0pt}
\begin{figure}[ht]
\setlength{\belowcaptionskip}{-10pt}
\begin{subfigure}{.5\textwidth}
  \centering
  \includegraphics[width=1.05\linewidth]{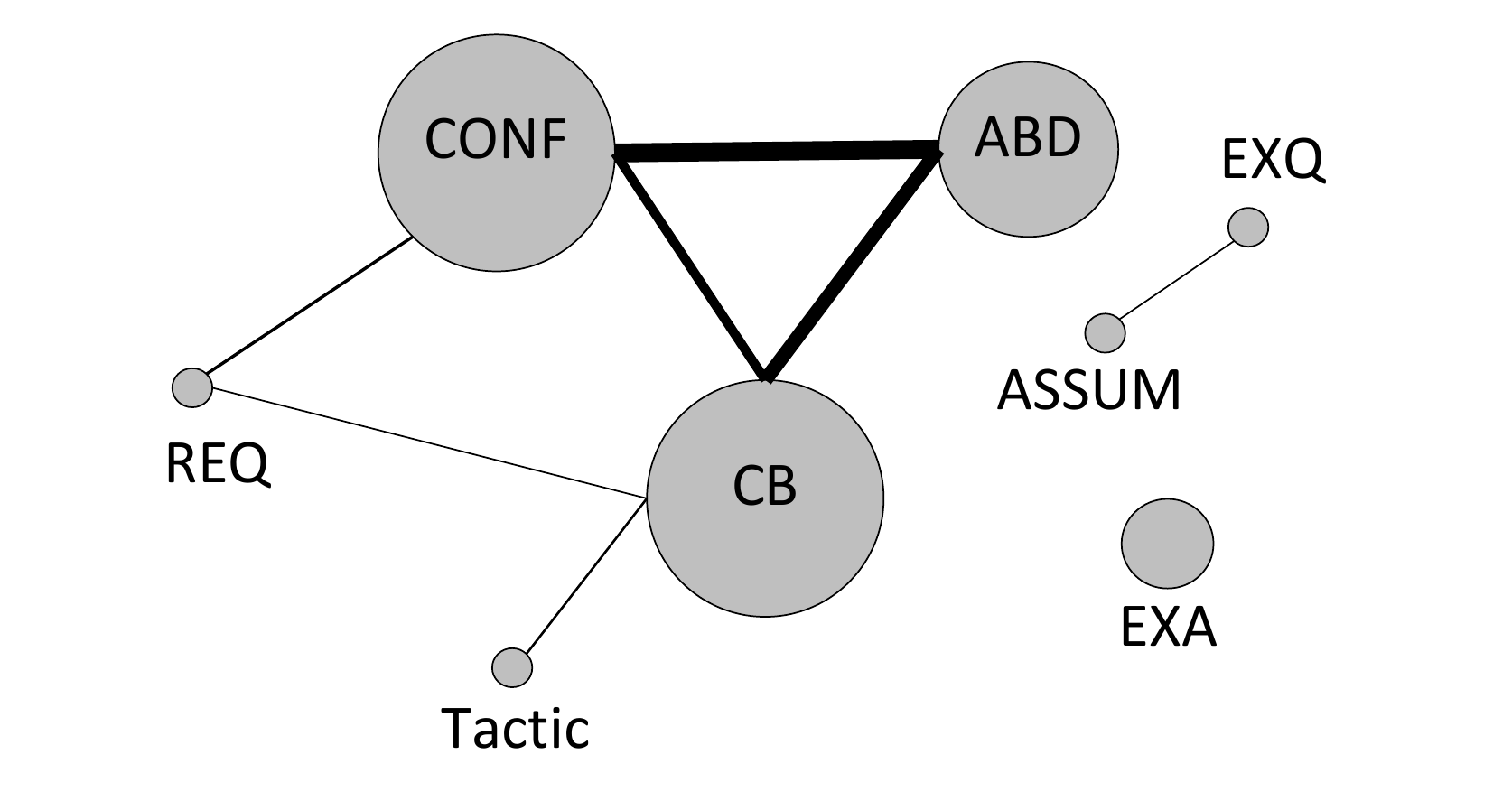}  
  \caption{ }
  \label{fig:sub-firstco}
\end{subfigure}
\begin{subfigure}{.5\textwidth}
  \centering
  \includegraphics[width=0.83\linewidth]{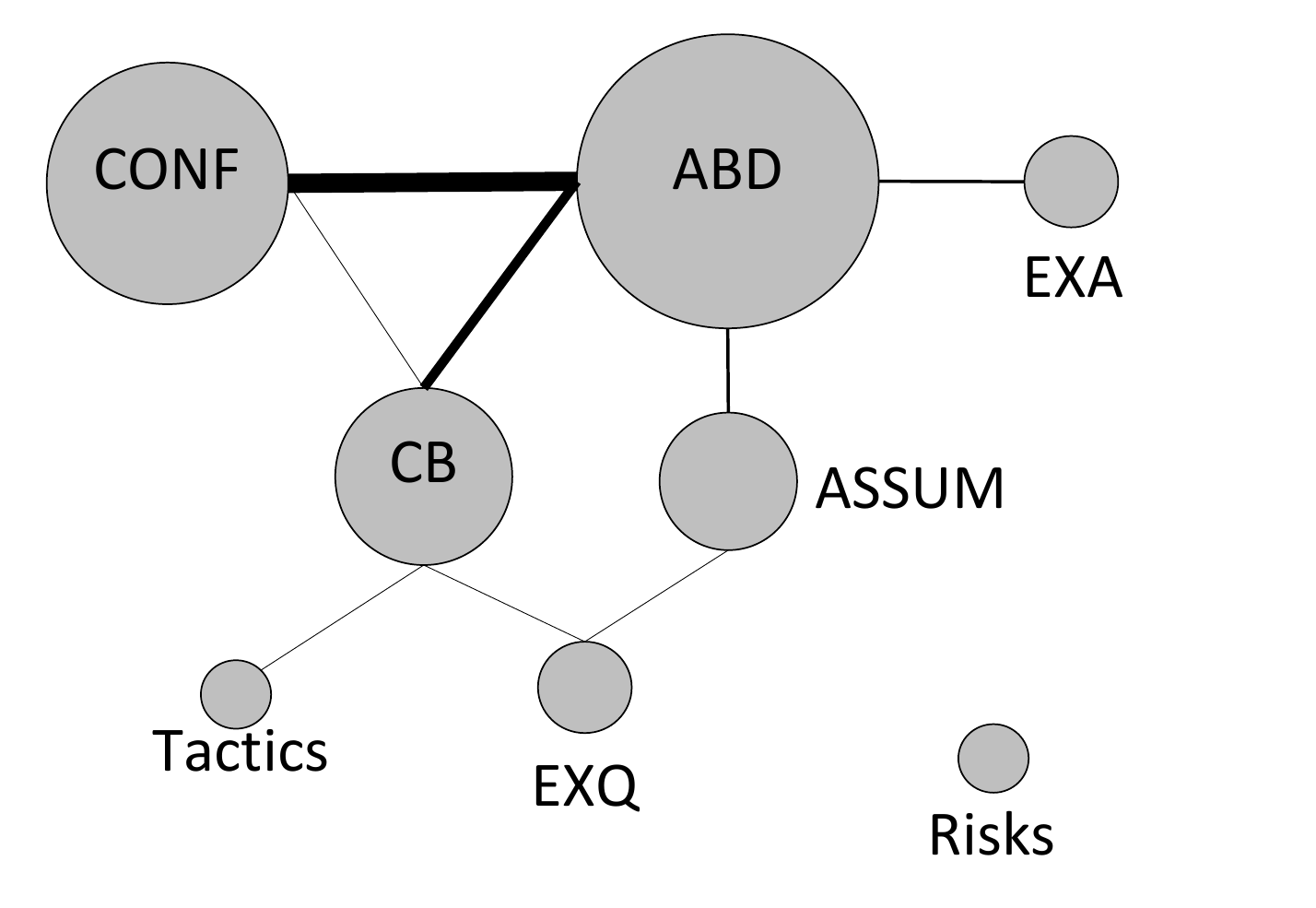}  
  \caption{ }
  \label{fig:sub-secondco}
\end{subfigure}
\caption{Significant co-occurrences (based $\tilde{\chi}^2$) between AK concepts in the description of issues (a) and issue comments (b). Node size indicates the significance of co-occurrences between annotations of the same AK concept with each other, while edge width is the significance of co-occurrence between annotations from different AK concepts.}
\label{fig:CoocurenceNetwork}
\end{figure}
\vspace{-5pt}
Fig. \ref{fig:CoocurenceNetwork} shows the co-occurrence networks for the significant co-occurrences between AK concepts in issue descriptions and comments, respectively. Significance has been computed using the $\tilde{\chi}^2$ test as explained in Section \ref{sec:studydesign}. From Fig. \ref{fig:CoocurenceNetwork}, we can observe that annotations of benefits and drawbacks (ABD) have the most significant co-occurrences with architectural configurations (CONF) and component behavior (CB), both in the issue description and comments. This means that practitioners usually share their AK on components design solutions, and associate it with their benefits and drawbacks as a rationale.

When comparing Fig. \ref{fig:sub-firstco} and \ref{fig:sub-secondco}, we observe some differences: First, requirements (REQ) (functional and non-functional) only significantly co-occur with other AK concepts in issue descriptions, but not in issue comments. This shows that discussions on architecturally significant requirements happen in issue descriptions rather than in issue comments. Second, assumptions (ASSUM) co-occur significantly only with Existing system quality (EXQ) (e.g. technical debt items) in issue descriptions, while they co-occur significantly with EXQ and Benefits and drawbacks (ABD) in issue comments. This shows that assumptions is a multifaceted AK concept, which appears in issue descriptions to express uncertainties about system quality, while in issue comments they additionally express uncertainty regarding the benefits and drawbacks of a proposed solution.
\vspace{-10pt}
\section{Discussion}
\label{sec:discussion}
\vspace{-5pt}
\subsection{RQ1: Representation of AK Concepts in Issues}
\subsubsection{Implications for practitioners}
The textual variants for AK concepts in Section \ref{sec:textualpatterns} can support practitioners to search for AK in issue trackers. For instance, to search for benefits and drawbacks, practitioners can use regular expressions, which require either adjectives and names of architectural solutions (e.g. tactics like caching) or quality attribute terms (e.g. scalability) in the same search. Moreover, the textual variants of AK concepts can provide ideas for documenting each AK concept. For example, following Table \ref{tab:rationaleAK}, practitioners could document benefits and drawbacks as follows: 1) create explicit lists of benefits and drawbacks, 2) describe the benefits and drawbacks of solutions using adjectives, 3) use quality measurement to justify the benefits and drawbacks, 4) mention explicitly problems of solutions as drawbacks.
\vspace{-10pt}
\subsubsection{Implications for researchers}
The textual variations for AK concepts in Section \ref{sec:textualpatterns} can support researchers to develop approaches to automatically extract AK concepts from issue trackers. Concretely, when using machine learning to identify and classify AK concepts from architectural issues, a corpus of annotations of AK concepts in issues is needed. A high quality corpus for improving the quality of classification would consider the different textual variants for AK concepts as presented in Section \ref{sec:textualpatterns}. Moreover, our corpus (3,937 annotations from 98 architectural issues) provides a starting point to train machine learning models on identifying and classifying AK concepts from issue trackers.

The variants of AK concepts in architectural issues show the benefits and challenges in capturing AK from issue trackers compared to developer communities (e.g. Stack Overflow). On the one hand, developers in issue trackers use domain-specific terms (e.g. ''optimizer'' or ''reduce'') to describe the architecture of a system, while developers in communities use generic terms (e.g. ''server'' or ''code''). This makes capturing generic AK from issue trackers more challenging. On the other hand, developers share AK in issues with extensive details; in contrast, developers in developer communities omit many of the details of a system when sharing their AK. 
This makes the AK in issue trackers more comprehensive than the AK in developer communities.
\vspace{-10pt}
\subsection{RQ2: AK Concepts in Issues}
\subsubsection{Implications for practitioners}
Knowing which AK concepts are mostly shared in architectural issues supports practitioners to effectively direct their search for AK. For instance, practitioners could search for AK in issue tracking system, if they are looking for alternatives of component design. This is because architectural solutions on component design present the majority of AK concepts in architectural issues. In contrast, it may not be effective to search in issue tracking systems for architecturally significant requirements (e.g. quality attributes), because these are rarely discussed in issue tracking systems.
\vspace{-10pt}
\subsubsection{Implications for researchers}
The results from RQ2 support researchers determining architectural scenarios, in which the re-use of AK from issue tracking system could be most useful. Specifically, the AK in architectural issues could be useful in these two scenarios:
\begin{itemize}
\item \textit{Architectural recovery}: Current architectural recovery techniques capture components and their dependencies from source code or byte code. However, such approaches cannot easily capture the behavior of components solely based on the source code. Thus, AK in issue tracking systems could complement architectural recovery techniques with additional natural language descriptions for component behaviors and configurations.
\item \textit{Selecting architectural solutions}: Results of RQ2 in Section \ref{sec:AKConceptsNumbers} shows that 40\% of AK concepts contain the rationale of ADDs (mostly benefits and drawbacks of solutions). Re-using this AK could facilitate comparing architectural solutions and selecting among alternatives based on their pros and cons.
\end{itemize}
\vspace{-9pt}
\subsection{RQ3: Significant Co-occurrences between AK Concepts}
\subsubsection{Implications for practitioners}
Because architectural issues involve lots of discussions, the significant co-occurrences between AK concepts in issues (as presented in Fig. \ref{fig:CoocurenceNetwork}) can guide practitioners when browsing for AK in architectural issues. For instance, if a practitioner found an architectural solution (e.g. an architectural configuration) in a comment, she may keep looking for the rationale of this solution: it is likely written in the same comment. The same applies when finding quality attributes or functional requirements in issue descriptions; these are likely to be accompanied with a certain architectural solution (i.e. an architectural configuration or component behavior).
\vspace{-10pt}
\subsubsection{Implications for researchers}
The significant co-occurrences between AK concepts can guide AK extraction approaches to effectively identify the relationships between AK concepts. For example, we can design a heuristic-based AK extraction approach, which links annotations on architectural configurations with annotations on benefits and drawbacks from the same issue section (i.e. issue description or comment); based on our results in Fig. \ref{fig:CoocurenceNetwork}, architectural configurations and benefits and drawbacks significantly co-occur in the same issue section. Associate architectural solutions with their rationale is very useful for the re-use of AK. 

Moreover, some significant co-occurrences between AK concepts are worth further detailed analysis. For example, Assumptions seem to co-occur with different AK concepts, especially decision factors like technical debt items, as well as rationale of decisions like benefits and drawbacks. However, it is unknown why and how such significant co-occurrences happen. Thus, further research could determine how and why assumptions co-occur with each of the AK concepts.
\vspace{-10pt}
\section{Threats to Validity}
\label{sec:threats}
\vspace{-5pt}
\subsection{External Validity}
Similar to other studies (\cite{Bhat2017AutomaticApproach,Shahbazian2018RecoveringDecisions}), our study depends on selecting issues from open source Apache projects and analyzing issues from Jira. This might be a threat, when generalizing the results to industrial projects or other ecosystems. Moreover, our analysis is based on a limited number of architectural issues and annotations, which might be a threat to the generalizability of our results. However, we have carefully selected these issues since they contain the most AK in the projects. Moreover, we created a significant sample of annotations, which are sufficient to report our quantitative results in Sections \ref{sec:AKConceptsNumbers} and \ref{sec:AKConceptsCooc} with the smallest  error margin as possible.
\vspace{-10pt}
\subsection{Construct Validity}
The considered AK concepts in Section \ref{sec:background} might not be exhaustive. However, during our qualitative analysis (see Section \ref{sec:studydesign}), we reached theoretical saturation, and thus covered most AK concepts in architectural issues.
\vspace{-10pt}
\subsection{Reliability}
The agreement on the AK concept for each annotation presents a threat to reliability. However, we considered the agreement between researchers in each phase of the study. Moreover, we created a coding book (provided online) with concrete textual variants to facilitate reaching agreement on the AK concepts. We measured the agreement between researchers using Kappa as 0.73. This shows good agreement beyond chance. In addition, we provide the list of identified architectural issues and annotations to support further replication steps.
\vspace{-9pt}
\section{Related Work}
\label{sec:related}
AK concepts (as presented in Section \ref{sec:background}) are established based on several studies, such as \cite{Kruchten2006,Tang2007a,Zimmermann2009}. Recent research efforts explore AK for specific domains (e.g. microservices \cite{ElMalki2019GuidingArchitectures}), as well as human and social aspects when making ADDs \cite{Razavian2019EmpiricalAnalysis}. However, these studies do not explore concrete representation of AK concepts in any AK source like issue trackers to support finding or capturing AK.

Recently, researchers explored and captured AK concepts in multiple different sources, such as developer communities (e.g. Stack Overflow \cite{Bi2021MiningOverflow,SolimanICSA2017,Wicsa2016,Soliman2018ImprovingCommunities}), Google search results \cite{Soliman2021ExploringKnowledge}, technology documentation \cite{GortonICSA2017} and mailing lists \cite{Fu2021WillDecisions}. However, these studies do not explore or capture AK in issue tracking systems.

Bhat et al. \cite{Bhat2017AutomaticApproach} propose a machine learning approach to classify issues in issue trackers, which contain certain types of ADDs. The approach depends on classifying issue descriptions without considering issue comments or attachments. Moreover, the approach does not explore different AK concepts (see Section \ref{sec:background}) in details. Our study on the other hand explores different AK concepts, their textual representation and relationships, considering the different sections of issues in more details.

Shahbazian et al. \cite{Shahbazian2018RecoveringDecisions} proposed an approach to identify architectural issues by analyzing source code and applying clustering algorithms. The approach can additionally differentiate between simple, compound and crosscutting decisions, and identify relationships between source code and issues. However, Shahbazian et al. did not analyze the textual content of architectural issues to explore the different AK concepts. The approach from Shahbazian et al. has inspired our study design to identify candidates architectural issues. However, our study goes beyond identifying architectural issues, and analyzes the textual content of architectural issues to explore the representation of AK concepts and their relationships within issues.
\vspace{-10pt}
\section{Conclusions}
\label{sec:conclusion}
\vspace{-5pt}
Our goal in this study is to explore the AK concepts within architectural issues to support re-using this AK. Our results cover existing AK concepts within architectural issues, as well as their textual representation and relationships. These support determining how the AK in issues could be re-used. Our future work aims to expand our study to explore different types of decisions in issue trackers, and to identify and extract the AK concepts automatically from issues.

\vspace{-10pt}
\bibliography{references}
\end{document}